\documentstyle[epsfig,psfig]{texas}

\begin{document}

\def\apj{ApJ}
\def\pf{Phys. Fluids}	
\def\prl{PRL}				
\def\keV{\hbox{keV}}
\def\MeV{\hbox{MeV}}
\def\Hz{\hbox{Hz}}
\def\cm{\hbox{cm}}
\def\G{\hbox{G}}
\def\gm{\hbox{gm}}

\title{A Plasma Instability Theory of Gamma-Ray Burst Emission}

\author{J. J. Brainerd}
\address{University of Alabama in Huntsville\\
E-Mail: Jim.Brainerd@msfc.nasa.gov
}

%
%
\begin{abstract}

A new theory for gamma-ray burst radiation is presented.
In this theory, magnetic fields and relativistic electrons are
created through plasma processes arising as a relativistic shell
passes through the interstellar medium.  The gamma-rays are
produced through synchrotron self-Compton emission.
It is found that shocks do not arise in this theory,
and that efficient gamma-ray emission only occurs for
a high Lorentz factor and a high-density interstellar medium.
The former explains the absence of gamma-ray bursts with thermal
spectra.  The latter provides
the Compton attenuation theory with an explanation of why
the interstellar medium density is always high.
The theory predicts the existence of a class of
extragalactic optical transient that emit no gamma-rays.
\end{abstract}

%
%

\section{Introduction}

A plasma instability theory for the prompt
emission of gamma-ray bursts is presented.\cite{brainerd4}
In this theory, a relativistic shell with $\Gamma \gg 1$
passes through the interstellar medium.  Two plasma
instabilities, the filamentation instability and
the two-stream instability, generate a magnetic field
and heat the electrons to relativistic energies.
The heated electrons emit synchrotron radiation in the
radio to optical bands and synchrotron
self-Compton radiation from the optical to gamma-ray
bands.

This theory produces the observed
prompt gamma-ray emission seen in all bursts, and
the prompt optical emission seen in
GRB~990123.  The magnetic field generated by the
filamentation instability is calculated from first
principals.  Lower limits on $\Gamma$ and $n_{ism}$
arise from the requirement that the model efficiently
radiate gamma-rays.  The limit on density requires
each gamma-ray burst to be surrounded by a medium that
is optically thick to Compton attenuation.
The limit on $\Gamma$ suggests that there exists a class
of transient that produces optical and ultraviolet emission
but no gamma-ray emission.

One of the more interesting aspects of the theory
is that the plasma instabilities cannot satisfy
the Rankine-Hugoniot conditions, so a shock is
not produced by these instabilities.  As
a consequence, the interstellar medium remains
in place after passage of the relativistic shell.
This permits the interstellar medium to interact with
multiple relativistic shells to produce
the complex time structure seen
in gamma-ray burst light profiles.

\section{Plasma Instabilities}

Two instabilities arise when a plasma streams through a second
plasma at a highly relativistic velocity.  The first is the
filamentation instability, while the second is the two-stream
instability.\cite{davidson1}
Of these, the former has the higher growth rate.

The growth rate of the filamentation instability as measured
in the shell rest frame is
\begin{equation} 
\gamma^{\prime}_{f} \approx {1 \over 2} \sqrt{ { 4 \pi e^2 \over m } }
   \, n_{ism}^{{1 \over 2}}
   \, ,
\end{equation}
where $n_{ism}$ is the number density of the interstellar medium in
the ISM rest frame, and $m$ is the mass of the filamenting plasma
component of the ISM. 
The filamentation occurs for wave numbers perpendicular to the
velocity vector that obey the inequality
\begin{equation}
k_{\perp} > \omega_{p,e,shell}^{\prime}/c \, .
\end{equation}
In other words, the length scale of the filamentation is set by the
electron plasma
frequency of the shell as measured in the rest frame of the shell.

\begin{figure}

\centering

\epsfig{figure=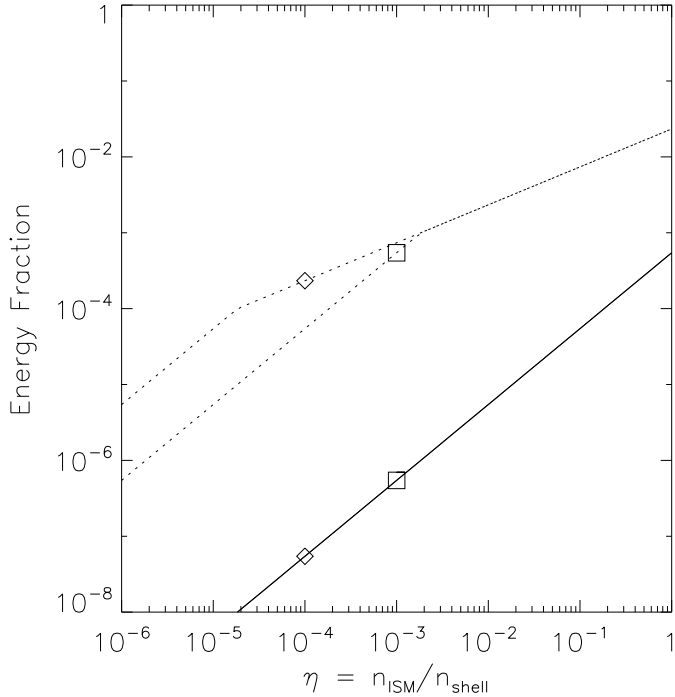, width=4.0in}
\caption{Fraction of energy that goes into magnetic field
and ion thermalization through the filamentation
instability.  This is measured in the rest frame of the
shell. The energy available is the energy of the interstellar medium
streaming through the shell.
The curves are plotted as the ratio $n_{ism}/n_{shell}^{\prime}$.
The solid curve gives the fraction of energy in magnetic field.  This
fraction is independent of $\Gamma$.  The dotted curve gives the fraction
of energy that goes into ion thermal energy.  The upper curve is for
$\Gamma = 10^4$, while the lower is for $\Gamma = 10^3$.  The squares
mark the value of $\eta$ that one expects from the Rankine-Hugoniot
relations for $\Gamma = 10^3$, while the diamonds mark this relation
for $\Gamma = 10^4$.
}
\end{figure}

The filament is a magnetic pinch, with a toroidal magnetic field that is
formed when the particles in the filament collapse to the center of the
filament.
The filament grows until its growth rate equals the bounce frequency of a
particle across the magnetic pinch.  This saturation
defines the maximum magnetic field that can be generated.\cite{davidson2,lee}
For both ions
and electrons, the maximum field strength is the same
\begin{equation}
{B^{\prime \, 2} \over 8 \pi }
   = { m_e c^2 n_{ism}^2 \Gamma^2 \over n_{shell}^{\prime} }
\end{equation}
One finds that for
$n_{shell}^{\prime}/n_{ism} = \Gamma$, which is a natural value
for the efficient emission of radiation,
the magnetic field is $B^{\prime} = 0.14 \, \G$ when $\Gamma = 10^3$ and
$n_{ism} = 1 \, \cm^{-3}$, and $B^{\prime} = 45.4 \, \G$
when $\Gamma = 10^3$ and $n_{ism} = 10^5 \, \cm^{-3}$.
The energy extracted from the interstellar medium and converted
into magnetic and thermal
energy is small.  This is shown in Figure~1.  As a consequence, even when
the Rankine-Hugoniot condition on $n_{shell}^{\prime}$ of 
$n_{shell}^{\prime}/n_{ism} = \Gamma$ is satisfied,
the fraction of kinetic energy that is converted to thermal and magnetic
energy is tiny, so that the interstellar medium continues to stream
through the shell.

\begin{figure}

\centering

\epsfig{figure=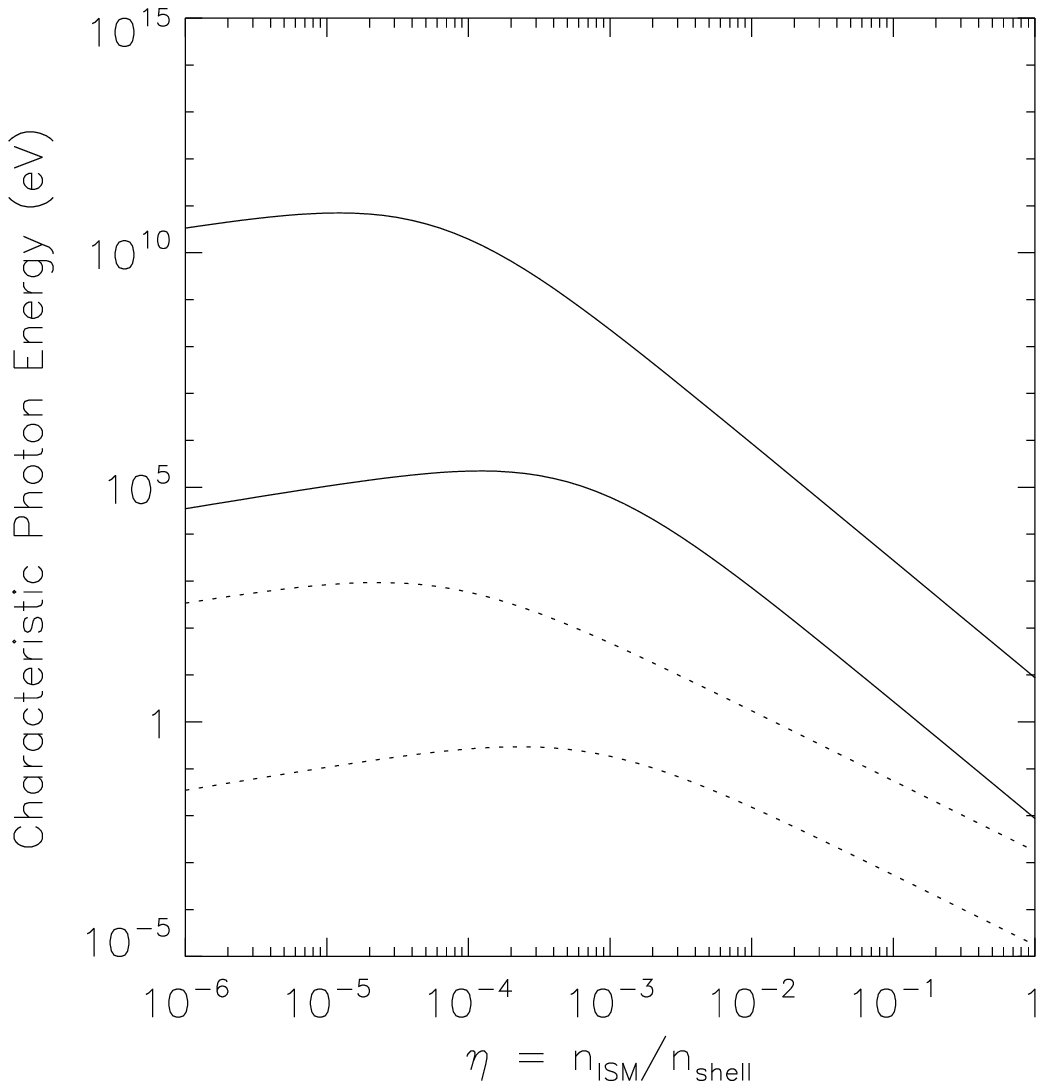, width=4.0in}
\caption{Characteristic observed
photon energies for $n_{ISM} = 1 \, \cm^{-3}$.
The solid curves give characteristic energies
for single scattering synchrotron self-Compton emission.
The dotted curves are for synchrotron emission.  The upper
pair are for $\Gamma = 10^4$, while the lower pair
are for $\Gamma = 10^3$.  Note that each of these curves
peak at $\eta \approx \Gamma^{-1}$.  Only the synchrotron
self-Compton emission is capable of producing gamma-ray
emission.
}
\end{figure}

\begin{figure}

\centering

\epsfig{figure=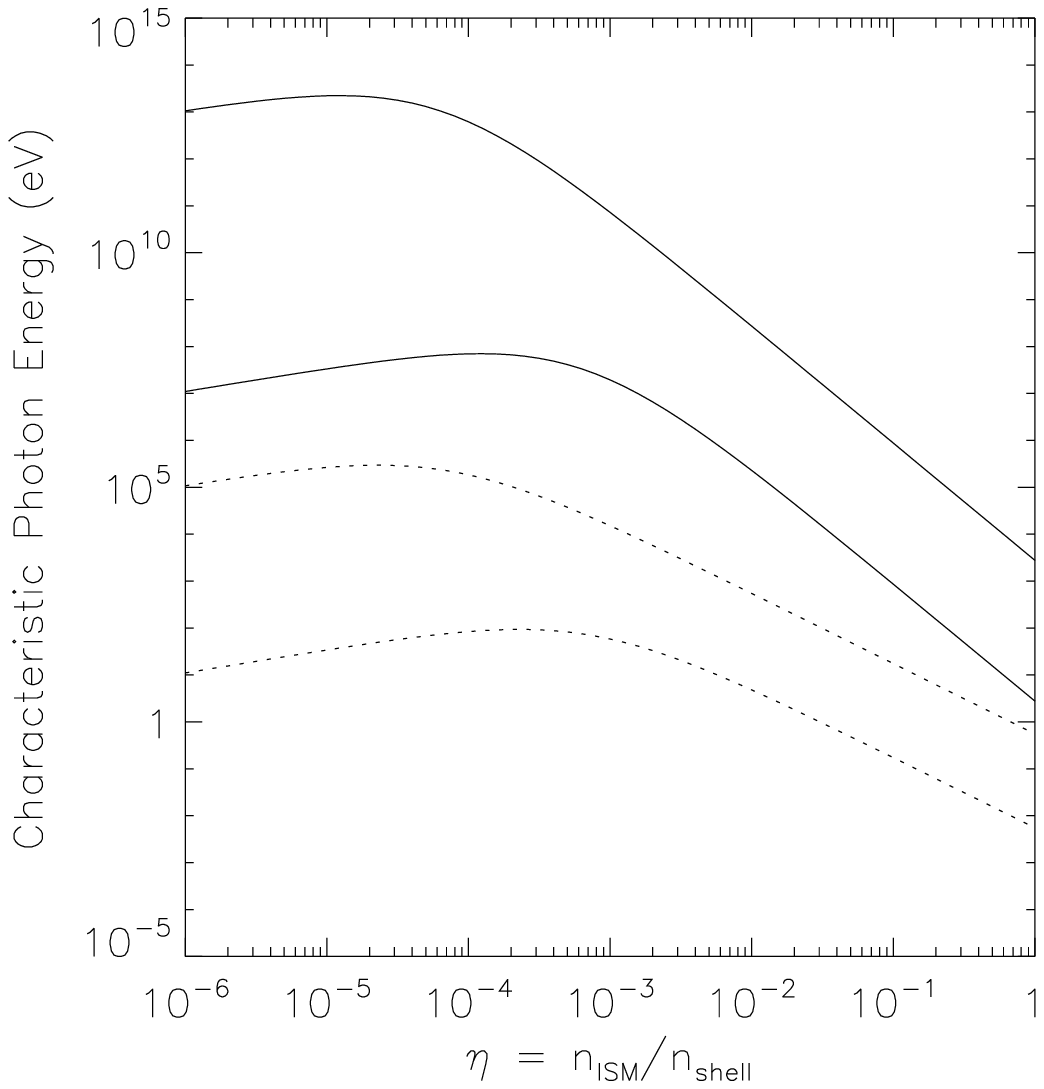, width=4.0in}
\caption{Characteristic observed
photon energies for $n_{ISM} = 10^{5} \, \cm^{-3}$.  These
curves are as in Figure 3a.  As in the previous case, only
the synchrotron
self-Compton emission is capable of producing gamma-ray
emission, although the synchrotron emission can produce
hard x-ray emission for case of $\Gamma = 10^4$.
}
\end{figure}

The two-stream instability acts on the electrons to bring them into
an equilibrium with the two ion streams.  This instability
grows at the rate of
\begin{equation}
\gamma^{\prime}_{2s} \approx {1 \over 2} \sqrt{{ 4 \pi e^2 \over m_e}}
n_{ism}^{{1 \over 2}} \Gamma^{-1} \, .
\end{equation}
The growth rate of the electron two-stream instability is
smaller than the ion filamentation instability by the factor of
$m_{p}^{1/2}/m_e^{1/2} \Gamma \approx 0.04$.  The two stream
instability will grow until the electron Lorentz factor is of order
$\Gamma$.  At this point, the system becomes stable.  This instability
therefore only converts $m_e/m_p$ of the total energy of the interstellar
medium into thermal energy.

\section{Radiative Mechanisms}

The two radiative mechanisms at work in this theory are synchrotron
emission and synchrotron self-Compton emission.  The characteristic
energy of each process is given in Figures 2 and 3 for
$n_{ism} = 1 \, \cm^{-3}$ and $n_{ism} = 10^5 \, \cm^{-3}$.  This
characteristic energy defines the maximum energy
of the continuum as defined by the maximum energy of the electron
distribution.  The minimum energy of the synchrotron emission
is given by the cyclotron energy, which is
\begin{equation} 
\nu = 2.68 \times 10^{8} \, \Hz \,
   n_{ism}^{{1 \over 2}} \Gamma_3^{{3 \over 2}} \, ,
\end{equation}
where $\Gamma_3 = \Gamma/10^3$.
This value is smaller than the characteristic synchrotron energy by
the factor of $\Gamma^2$.

The synchrotron self-Compton energy range is determined by a single
scattering, since more than one scattering takes the photons to the
characteristic electron energy.
The minimum energy is set by $\Gamma^2$ times the cyclotron resonance
energy, so that the low end of the synchrotron self-Compton continuum
overlaps the high end of the synchrotron continuum.
The characteristic Compton scattering
energy is a factor of $\Gamma^2$ larger than this.  From the figures,
one sees that for $\Gamma > 10^3$, the energy range of the Compton scattered
radiation extends above $1 \MeV$.
For a lower value of $\Gamma$, the cooling occurs predominately at
optical and ultraviolet wavelengths.  Because gamma-ray bursts are
identified by their gamma-ray emission, bursts with $\Gamma < 10^3$ will
not be observed.  This suggests that there exists a class of burst
phenomena with optical and ultraviolet emission, but no gamma-ray emission.

\section{Selection Effects}

The emission of gamma-rays defines one selection effect.  In order to
produce gamma-rays, $\Gamma > 10^3$.  This provides an explanation
for the absence of photon-photon pair creation in gamma-ray bursts,
as demonstrated by the absence of gamma-ray bursts with thermal
spectra.

\begin{figure}

\centering

\epsfig{figure=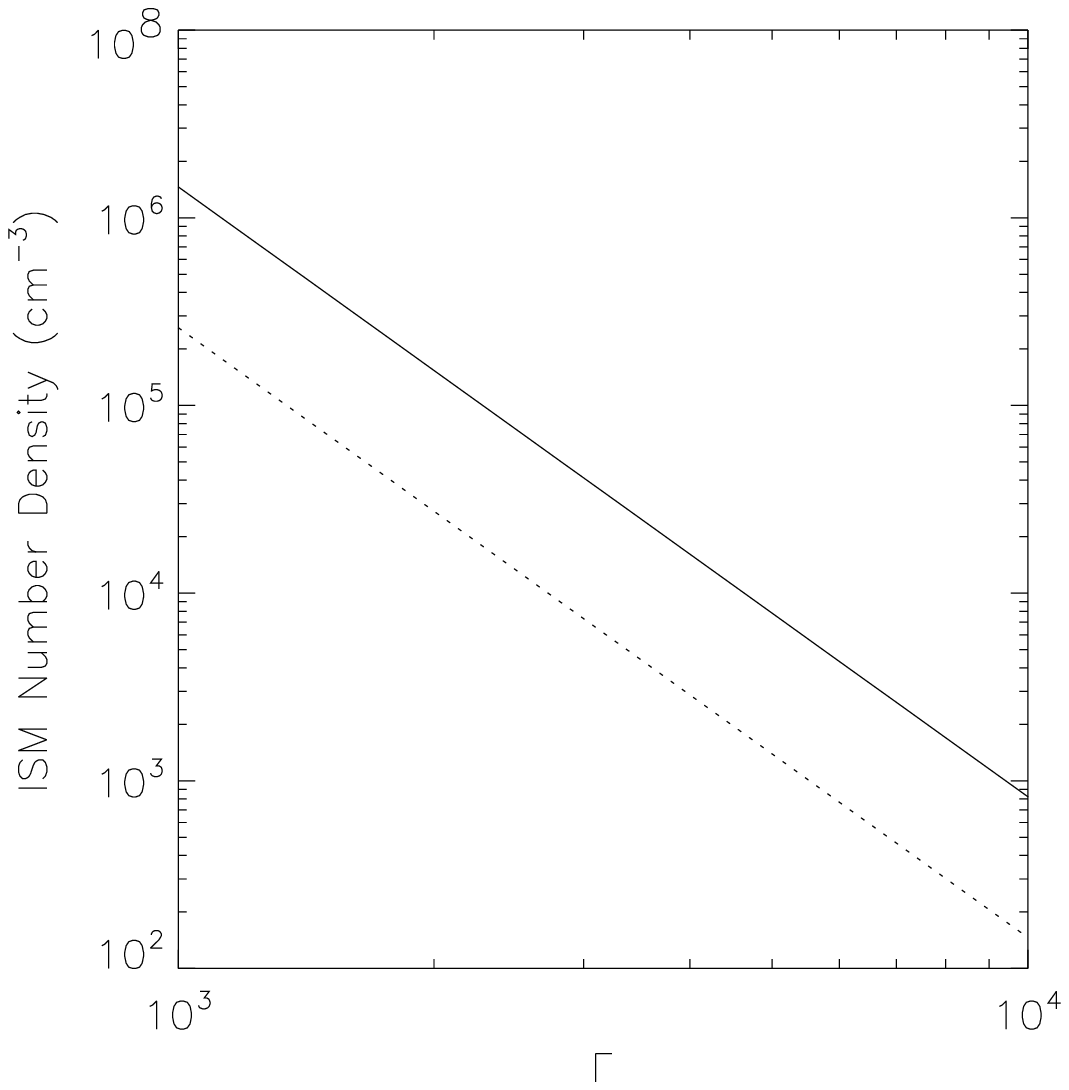, width=4.0in}
\caption{Lower limits on the interstellar medium
number density.  This is as a function of the shell Lorentz
factor $\Gamma$.  The solid curve is for radiative processes
radiating at a rate equal to the rate at which energy is transferred
to electrons from the streaming ions.  The dotted curve is for radiative
processes radiating at 10\% of the rate at which energy is
transferred to electrons.  To efficiently radiate energy, the interstellar
medium number density must be above these curves.  The provides a
selection effect: only burst sources surrounded by high density material
will produce observable gamma-ray bursts.
}
\end{figure}

A second selection effect that comes into play in this theory
concerns the efficient production of gamma-rays.  If the density is
too low, the rate of radiative cooling is far below the rate at which
electrons are thermalized.
Such bursts would be dim relative to their high-density
counterparts.  The most observable bursts are
therefore those with a sufficiently high $n_{ism}$ to
efficiently convert electron thermal energy into gamma-rays.
This lower limit on $n_{ism}$ can be written as
\begin{equation}
n_{ism} > 8.22 \times 10^{4} \cm^{-3}
   {\cal M}_{27}^{-{1 \over 2}} \Gamma_3^{-{13 \over 4}}
   f_{emis}^{-{3 \over 2}} \left( { R \over R_0 } \right)^{{3 \over 4}}
\end{equation}
where $f_{emis}$ is the fraction of electrons at the characteristic energy,
where $R/R_0 \approx \left( m_p/m_e \right)^{1/3}$ is the distance traveled
when $\Gamma$ drops by a factor of 2 relative
to the distance traveled when the ISM is swept up, all under the assumption that
the heating of the electrons represent the maximum energy loss,
and where ${\cal M}_{27}$ is the mass of the shell per unit ster radian
in units of $10^{27} \, \gm$.

The limit given by this last equation is plotted in Figure 4 as a solid
line.  A similar curve that gives emission that is 10\% effective
at radiating the energy converted into thermal energy is plotted as
a dotted line.  The importance of this limit is that the gamma-ray
burst mechanism requires a high density interstellar medium to operate
efficiently.  When the burst radiates efficiently, the interstellar
medium is of sufficient density to attenuate the
gamma-ray spectrum through Compton scattering.\cite{brainerd1}
This provides the plasma instability theory with a mechanism that gives
the spectrum a characteristic energy of
several hundred keV, despite the broad spectral
range of the synchrotron self-Compton
continuum, and its strong dependence on $\Gamma$.
Such a mechanism is required to correctly reproduce
the observed spectra.\cite{brainerd2,brainerd3}

%
%

\section{Discussion}

The plasma instability theory discussed above is a new mechanism to
produce prompt gamma-ray burst emission.  It has several unique features.

\begin{itemize}

\item The plasma instability theory produces
the prompt gamma-ray emission without creating a shock.  This implies
that additional gamma-ray bursts can occur in the
interstellar medium when new relativistic shells are ejected by
the source, because the region is not cleared of material.

\item The requirement that the mechanism efficiently
produce gamma-rays introduces selection effects, so that
bursts have $\Gamma > 10^3$ and $n_{ism} > 10^5 \cm^{-3}$.

\item There exists a class of optical transient that has no gamma-ray
emission.  These bursts differ from gamma-ray bursts in having $\Gamma < 10^3$.

\item The lower limit on $n_{ism}$ ensures that gamma-ray bursts
are always in region in which Compton attenuation by the surrounding
interstellar medium occurs.

\end{itemize}

The next step in developing this theory is to undertake a numerical
study of the plasma instabilities.  The goal of the study will be
to confirm the analytic results discussed above, and to provide
a precise calculation of the electron distribution produced by
the instabilities.  A second aspect of the theory that will be
examined is the afterglow radiation produced in this theory;
the cooling of the region behind the relativistic shells gives
a light curve that differs from the light curve created by
the decelerating shell, adding a complexity to the afterglow
from this theory that is not present in the shock theories
of afterglow radiation.  Finally, calculations of model spectra
and their comparison to observed spectra is will be undertaken.
In particular, a comparison of the optical and gamma-ray
spectrum of this model to that of GRB~990123\cite{akerlof}
is planned.  This study will test whether the prompt optical
and gamma-ray emission are part of a single synchrotron
self-Compton continuum.


\section*{References}


\begin{thebibliography}{99}

\bibitem{akerlof}
Akerlof, C. W., \& McKay, T. A. 1999, GCN Circ. 205

\bibitem{brainerd1}
Brainerd, J. J. 1994, \apj, 428, 21

\bibitem{brainerd2}
Brainerd, J.~J. \ 1996, in {\it Gamma-ray Bursts: Third Symposium,
Huntsville, AL 1995\/,} ed. C. Kouveliotou, M. S. Briggs, \& G.~J. Fishman
(New York: AIP), 148

\bibitem{brainerd3}
Brainerd, J. J., Preece, R. D., Briggs, M. S., Pendleton, G. N.,
\& Paciesas, W. S. 1998, \apj, 501, 325

\bibitem{brainerd4}
Brainerd, J. J. 1999, \apj, submitted

\bibitem{davidson1}
Davidson, R. C. 1990, ``Physics of Nonneutral Plasmas''
(Redwood City: Addison-Wesley)

\bibitem{davidson2}
Davidson, R. C., Hammer, D. A., Haber, I., \& Wagner, C. E.
1972, \pf, 15, 317

\bibitem{lee}
Lee, R., \& Lampe, M. 1973, \prl, 31, 1390

\end{thebibliography}
\end{document}